\documentclass[aip,jcp,preprint,a4paper,11pt]{revtex4-1}

\usepackage{amssymb}
 \usepackage{amsmath}
\usepackage{wasysym}
\usepackage{longtable}
\usepackage{dcolumn}
\usepackage{graphicx}
\usepackage{longtable}
\usepackage[usenames,dvipsnames]{color}
\newcommand{\marked}[1]{}

\newcommand{\mathnewydd}[1]{#1}
\newcommand{\newydd}[1]{#1}

\newcommand{\qq}[1]{{\lq}#1{\rq}}

\begin{document}

\date{4th August 2016}
\title{Molecular simulation of the surface tension of real fluids}

\author{Stephan Werth}
\author{Martin Horsch} 
\thanks{Author to whom correspondence should be addressed: martin.horsch@mv.uni-kl.de}
\author{Hans Hasse}
\affiliation{Laboratory of Engineering Thermodynamics, University of Kaiserslautern, Erwin-Schr\"odinger-Str. 44, 67663 Kaiserslautern, Germany}

\begin{abstract} 
Molecular models of real fluids are validated by comparing the vapor-liquid surface tension from molecular dynamics (MD) simulation to correlations of experimental data.
The considered molecular models consist of up to 28 interaction sites, including Lennard-Jones sites, point charges, dipoles and quadrupoles. They represent 38 real fluids, such as ethylene oxide, sulfur dioxide, phosgene, benzene, ammonia, formaldehyde, methanol and water, and were adjusted to reproduce the saturated liquid density, vapor pressure and enthalpy of vaporization. The models were not adjusted to interfacial properties, however, so that the present MD simulations are a test of model predictions. 
It is found that all of the considered models overestimate the surface tension. In most cases, however, the relative deviation between the simulation results and correlations to experimental data is smaller than 20 \%. This observation corroborates the outcome of our previous studies on the surface tension of 2CLJQ and 2CLJD fluids where an overestimation of the order of 10 to 20 \% was found.
\end{abstract}


\maketitle

\section{Introduction}

Interfacial properties are important for many applications in process engineering, including processes like absorption, wetting, nucleation, cavitation or foaming. Experimental data on the surface tension are available for pure fluids, but the temperature range is usually limited to ambient conditions \cite{DIPPR, DDB}. Hence, it is desirable to have models which allow predicting interfacial properties of pure fluids and mixtures over a wide temperature and pressure range. Molecular modelling and simulation can be used for this purpose if the underlying force fields are accurate \cite{GMT16}.

In previous work of our group \cite{DMVH12,MVH12b,MVH12,EVH08a,EVH08b,HHHV11,EHVH07,EVH08c,SCVLH07,SSVH07,SVH08,HMHHV12,VSH01} and recent work by Vrabec and co-workers \cite{MGSV13,EWSV12,EJKMMV14,KNWRV12,TDRWKSV16,TRKDWSV16} a large number of molecular models for real fluids were developed. The molecular model parameters were adjusted to describe the saturated liquid density, vapor pressure and enthalpy of vaporization, which they do well. These models were also used to predict transport properties and they showed very good agreement for the shear viscosity, self diffusion coefficients and thermal conductivity of pure fluids \cite{GVH12a,MVH12,EWSV12,TDRWKSV16,TRKDWSV16} and mixtures \cite{GVH12a,PGHV13,GVH11,GVH12,GJMV16}.
Some of the fluids discussed in the present work have recently been used to develop fundamental equation of state based on molecular simulation as well as experimental data, e.g.\ ethylene oxide \cite{TRKKSV15}, phosgene \cite{RV15}, hexamethyldisiloxane \cite{TDRWKSV16} and octamethylcyclotetrasiloxane \cite{TRKDWSV16}. The surface tension was not part of the parameterization and is thus strictly predictive.

In previous work systematic evaluations of the surface tension of the two center Lennard-Jones plus point quadrupole (2CLJQ) and the two center Lennard-Jones plus point dipole (2CLJD) molecular model class were conducted \cite{WHH15a,WHH16}. These models, on average, overestimate surface tension by about 20 \% and 12 \%, respectively \cite{WHH16, WSKKHH15}. Other molecular models which have been adjusted to bulk properties, but not to interfacial properties, exhibit similar deviations \cite{NWLM11,EMRV13,ALGAJM11,HTM15,FLPMM11,EV15,SSDS09,ZLCMSA13,CMHHCS12}.\marked{~(Removed: \qq{Attempts to reconciliate the quality of the description of the bulk properties and that of the interfacial properties by including the latter into the fit of the parameters of the molecular models have so far not been successful \cite{SKRHKH14,WSKKHH15,SKHKH16,WSHH16}.})~}

In the present work,\marked{~(removed: \qq{the})~} existing molecular models are used straightforwardly. No parameters are changed. \newydd{By molecular dynamics (MD) simulation,} predictions of interfacial properties from bulk properties \newydd{are obtained. Used in this way, molecular modeling can be compared to other approaches for predicting the surface tension from bulk data, such as phenomenological parachor correlations \cite{Sugden24, HV86, GDDR89, ZS97}, corresponding-states or critical-scaling expressions \cite{ZS97, SR95}, which are also phenomenological correlations, and other molecular methods, e.g.~square gradient theory \cite{GMPBM16} and density functional theory \cite{Gross09, TG10} on the basis of molecular equations of state \cite{GS01, GV06, VG08, ALAGMJ13}.}

In the present work, bulk and interfacial properties of real fluids are determined simultaneously from heterogeneous \newydd{MD}\marked{~(removed: \qq{molecular dynamics})~} simulations. The simulation results are compared with correlations to experimental data, where available.

\section{Molecular simulation}

The molecular models discussed in the present work are taken from previous work of our group \cite{DMVH12,MVH12b,MVH12,EVH08a,EVH08b,HHHV11,EHVH07,EVH08c,SCVLH07,SSVH07,SVH08,HMHHV12,VSH01} and recent work by Vrabec and co-workers \cite{MGSV13,EWSV12,EJKMMV14,KNWRV12,TDRWKSV16,TRKDWSV16}.
The molecular models are internally rigid and consist of several Lennard-Jones sites and superimposed electrostatics. The total potential energy is given by
\begin{align}
 U = &\sum_{i=1}^{N-1} \sum_{j=i+1}^{N} \Bigg\{ \sum_{a=1}^{n_i^{\rm LJ}} \sum_{b=1}^{n_j^{\rm LJ}} 4 \epsilon_{ijab} \left[ \left( \frac{\sigma_{ijab}}{r_{ijab}} \right)^{12} - \left( \frac{\sigma_{ijab}}{r_{ijab}} \right)^{6} \right] + \sum_{c=1}^{n_i^{e}} \sum_{d=1}^{n_j^{e}} \frac{1}{4\pi \epsilon_0} \Bigg[ \frac{q_{ic}q_{jd}}{r_{ijcd}} \notag \\
 &+ \frac{\mu_{ic}\mu_{jd}}{r_{ijcd}^3} \cdot f_1 ( \omega_i,\omega_j)+ \frac{\mu_{ic}Q_{jd}+\mu_{jd}Q_{ic}}{r_{ijcd}^4} \cdot f_2 ( \omega_i,\omega_j) + \frac{Q_{ic}Q_{jd}}{r_{ijcd}^5} \cdot f_3 ( \omega_i,\omega_j) \Bigg] \Bigg\},
\end{align}
where $\epsilon_{ijab}$ and $\sigma_{ijab}$ are the Lennard-Jones energy and size parameters, $r_{ijab}$ and $r_{ijcd}$ are site-site distances, $q_{ic}$, $q_{jd}$, $\mu_{ic}$, $\mu_{jd}$, $Q_{ic}$ and $Q_{jd}$ are the magnitude of the electrostatic interactions, i.e.\ the point charges, dipole and quadrupole moments, and $f_{k} (\omega_i,\omega_j)$ are dimensionless angle-dependent expressions in terms of the orientation $\omega_i$,$\omega_j$ of the point multipoles \cite{GG84}. 

\marked{\newydd{The following paragraph was moved here from the Introduction (and shortened):}}

Thermodynamic properties in heterogeneous systems are very sensitive to a truncation of the intermolecular potential \cite{ZLCMSA13,GMT14,WHVH15,GMT15,WHH15,GMSBRF04}. For dispersive interactions, like the Lennard-Jones potential, various long range correction (LRC) approaches exist which are known to be accurate for planar fluid interfaces \cite{MWF97,MENT03,Janecek06,VIG07,IHMI12,TSBI14,WRVHH14}\newydd{.}\marked{~(Removed: \qq{\dots, ranging from Ewald summation techniques \cite{VIG07,IHMI12}, the Fast Multipole Method (FMM) \cite{MENT03} and Multilevel Summation (MLS) \cite{TSBI14} to})~}\newydd{ The simulations in the present work use} slab-based LRC techniques based on the density profile \cite{MWF97,Janecek06,WRVHH14}.\marked{~(Removed: \qq{Ewald summation techniques have the disadvantage that even highly optimized codes scale with $\mathcal{O} (N^{3/2})$ in terms of the number of molecules $N$ \cite{PPD88,ms2} and more sophisticated methods like PPPM Ewald scale with $\mathcal{O} (N \log N)$ \cite{HE88,IHMI12}. 
Methods like FMM, MLS and slab-based LRCs have a more favorable linear scaling, i.e.\ $\mathcal{O} (N)$ \cite{AFHLBDHKGHIHPPS13,MENT03,TSBI14,WRVHH14,HWXSS16}. In terms of the thermodynamic results, the different methods deliver a similar degree of accuracy for Lennard-Jones systems \cite{Janecek06,IHMI12,TSBI14,WRVHH14}.}~)} For polar interactions, LRCs based on Ewald summation are typically used for the simulation of vapor-liquid interfaces \cite{HE88,PPD88,EKMW04,IHMI12,MSG15}. However, a computationally efficient slab-based LRC based on the density profile can also be used for polar molecular models \cite{WHH15}. In terms of the thermodynamic results, the different methods deliver a similar degree of accuracy for the two-center Lennard-Jones plus point dipole fluid \cite{WHH16,MSG15,EKMW04}. \newydd{Therefore, the slab-based LRC technique described in previous work\cite{WHH15} is employed here both for dipolar electrostatic interactions and for dispersion.}

For the present series of MD simulations, systems were considered where the vapor and liquid phases coexist with each other in direct contact, employing periodic boundary conditions, so that there are two vapor-liquid interfaces which are oriented perpendicular to the $y$ axis. The interfacial tension was computed from the deviation between the normal and the tangential diagonal components of the overall pressure tensor \cite{WTRH83,IK49}, i.e.\ the mechanical route,
\begin{equation}
 \gamma = \frac{1}{2}\int_{-\infty}^\infty \text{d}y \left(p_\text{N} - p_\text{T} \right).
\end{equation}
Thereby, the normal pressure $p_\text{N}$ is given by the $y$ component of the diagonal of the pressure tensor, and the tangential pressure $p_\text{T}$ was determined by averaging over $x$ and $z$ components of the diagonal of the pressure tensor. The simulations were performed with the \newydd{MD}\marked{~(removed: \qq{molecular dynamics})~} code \textit{ls1 mardyn}~\cite{ls12013} in the canonical ensemble with $N$ = 16,000 particles. Further details on the MD simulations are given in the Appendix.

\section{Results and Discussion}

Tab.~\ref{NCLJX_tab:PredRealFluids} gives an overview of the molecular models investigated in the present work. All studied models are rigid, i.e.\ internal degrees of freedom are no accounted for. The deviations $\delta \rho'$ and $\delta p^{\rm S}$, which are reported for the models in Tab.~\ref{NCLJX_tab:PredRealFluids}, are taken from the corresponding publications and represent relative mean deviations of the simulated values from correlations to experimental data \cite{DMVH12,MVH12b,MVH12,EVH08a,EVH08b,HHHV11,MGSV13,EHVH07,EVH08c,SCVLH07,SSVH07,SVH08,HMHHV12,EWSV12,EJKMMV14,KNWRV12,TDRWKSV16,TRKDWSV16,VSH01}. The molecular simulations in previous work were performed with the Grand Equilibrium method \cite{VH02}. 

\begin{figure}[t]
\centering
 \includegraphics[width=8.25cm]{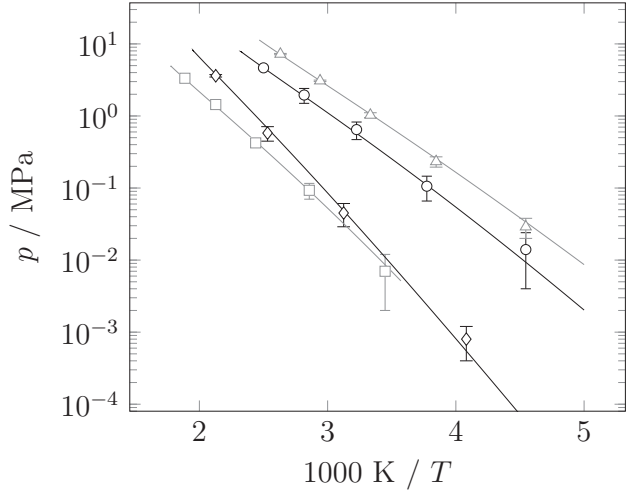}
\caption{Vapor pressure curves of ammonia, methanol, sulfur dioxide and benzene. Solid lines represent correlations to experimental data \cite{LS06,PPM92,THB93,RC93} and symbols are the present simulation results: Sulfur dioxide ($\circ$), benzene ($\square$), ammonia ($\triangle$) and methanol ($\diamond$).}  
\label{NCLJX_fig:pT}
\end{figure}

\marked{(Removed: \qq{Figs.~[previously 1] and \ref{NCLJX_fig:pT}})~}\newydd{Fig.\ \ref{NCLJX_fig:pT}} shows the simulation results for \newydd{the vapor pressure of} ammonia, methanol, sulfur dioxide and benzene from the present work which were obtained from heterogeneous simulations\newydd{; cf.\ the Appendix (Tab.~\ref{NCLJX_tab:data}) for a complete presentation of the simulated properties.} The \newydd{present}\marked{~(removed: \qq{simulation})~} results for the saturated densities and the vapor pressure are in very good agreement with experimental data \cite{LS06,PPM92,RC93,THB93}. However, for low temperatures the uncertainties in the vapor density and vapor pressure are relatively high. This is due to the fact that a low temperatures in many cases on average less than one molecule is in the vapor phase, which yields relatively high statistical uncertainties. Similar findings were obtained for the the other fluids studied in the present work. The simulation results obtained for the vapor pressure and the saturated densities for all studied fluids are reported in Tab.~\ref{NCLJX_tab:data} together with the data for the surface tension.


\marked{
   (Removed a Figure, i.e.\ originally Fig.\ 1, from the manuscript.)
}

The relative mean deviation $\delta\gamma$ between the simulation data and the experimental data reported in Tab.~\ref{NCLJX_tab:PredRealFluids} is calculated in the same way as the deviations for the saturated liquid density and the vapor pressure. It represents the relative mean deviation of the\marked{~(removed: \qq{predicted})~} surface tension \newydd{predicted} by the molecular models from Design Institute for Physical Properties (DIPPR) correlations to experimental data
\begin{equation}
  \mathnewydd{\left|\delta \gamma\right|} = \sqrt{\frac{1}{K} \sum_i^K \left( \frac{\gamma^{\rm\mathnewydd{sim}}(T_i)-\gamma^{\rm\mathnewydd{exp}}(T_i)}{\gamma^{\rm exp}(T_i)}\right)^2},
 \label{NCLJX_eq:devGamma}
\end{equation}
between \newydd{the} triple point temperature and 95 \% of the critical temperature.
\newydd{By convention, the sign of $\delta\gamma$ is positive if, on average, the model overestimates the surface tension ($\delta\gamma = + \left|\delta \gamma\right|$) and negative otherwise ($\delta\gamma = - \left|\delta \gamma\right|$).}
Underlying \newydd{experimental} surface tension data are usually not available over the entire temperature range. Only for four compounds - water, methanol, ammonia, heptafluoropropane - experimental data are available over the entire temperature range. In most cases, the surface tension is measured only up to 373 K and the DIPPR correlation\marked{~(removed: \qq{is then extrapolated towards})~} \newydd{extrapolates these results to} the critical point \cite{DIPPR,DDB}.

\begin{figure}[t]
\centering
 \includegraphics[width=8.25cm]{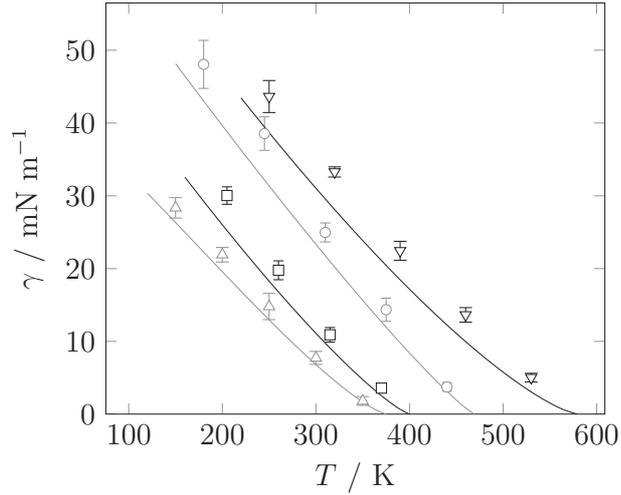}
\caption{Surface tension as a function of the temperature. Solid lines represent DIPPR correlations to experimental data \cite{DIPPR} and symbols are the present simulation results: Thiophene ($\triangledown$), ethylene oxide ($\circ$), dimethyl ether ($\square$) and heptafluoropropane ($\triangle$).}  
\label{NCLJX_fig:NCLJX1}
\end{figure}

\begin{figure}[h]
\centering
 \includegraphics[width=8.25cm]{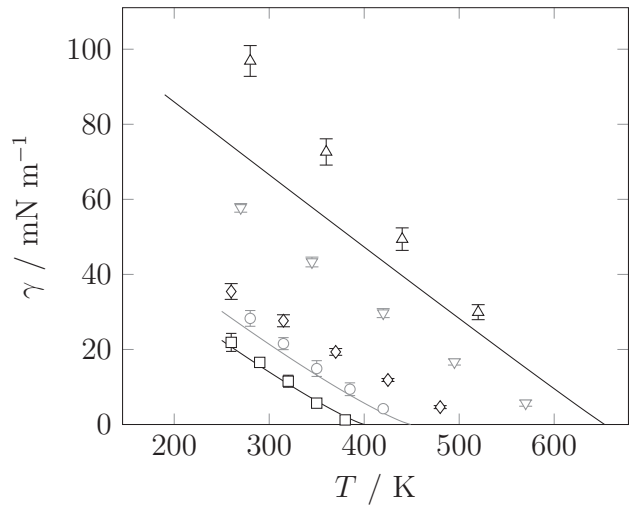}
\caption{Surface tension as a function of the temperature. Solid lines represent DIPPR correlations to experimental data \cite{DIPPR} and symbols are the present simulation results: Hydrazine ($\triangle$), methylhydrazine ($\triangledown$), 1,1-dimethylhydrazine ($\diamond$), cyanogen chloride ($\circ$) and cyanogen ($\square$). No experimental data are available for methylhydrazine and 1,1-dimethylhydrazine.}  
\label{NCLJX_fig:NCLJX2}
\end{figure}

The DIPPR correlations usually agree with available experimental data within 3 \%, only for dimethyl sulfide, ortho-dichlorobenzene, heptafluoropropane, cyanogen, decafluorobutane and hexamethyldisiloxane deviations of up to 5 \% are reported \cite{DIPPR}. For three fluids - formaldehyde, methylhydrazine and 1,1-dimethylhydrazine - no experimental data are available. The DIPPR correlations do not match the critical temperature for ethylene glycol and formic acid. Therefore, a straight line is used to connect the DIPPR correlation and the critical point of respective fluids.

\begin{figure}[h]
\centering
 \includegraphics[width=8.25cm]{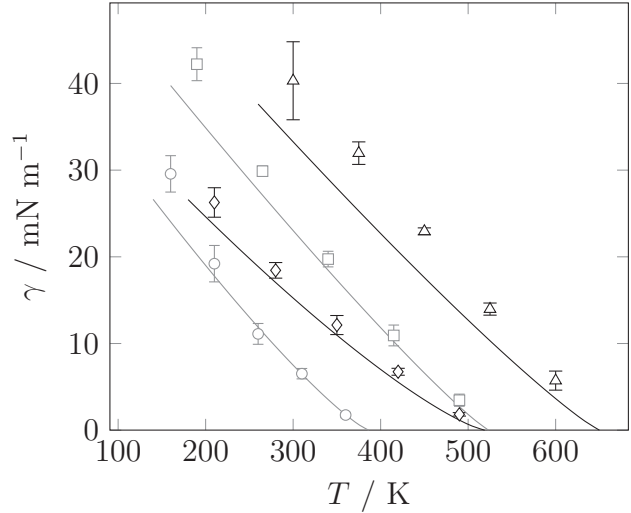}
\caption{Surface tension as a function of the temperature. Solid lines represent DIPPR correlations to experimental data \cite{DIPPR} and symbols are the present simulation results: Cyclohexanol ($\triangle$), ethyl acetate ($\square$), hexamethyldisiloxane ($\diamond$) and decafluorobutane ($\circ$).}  
\label{NCLJX_fig:NCLJX3}
\end{figure}

\begin{figure}[b]
\centering
 \includegraphics[width=8.25cm]{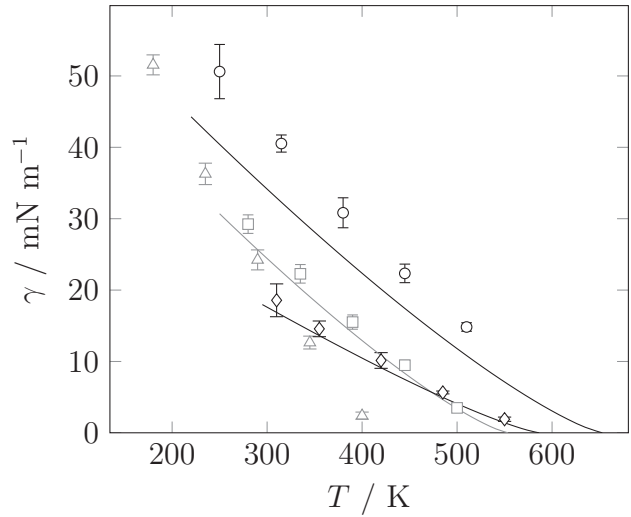}
\caption{Surface tension as a function of the temperature. Solid lines represent DIPPR correlations to experimental data \cite{DIPPR} and symbols are the present simulation results: Cyclohexanone ($\circ$), cyclohexane ($\square$), formaldehyde ($\triangle$) and octamethylcyclotetrasiloxane ($\diamond$). No experimental data are available for formaldehyde.}  
\label{NCLJX_fig:NCLJX4}
\end{figure}

\begin{figure}[b]
\centering
 \includegraphics[width=8.25cm]{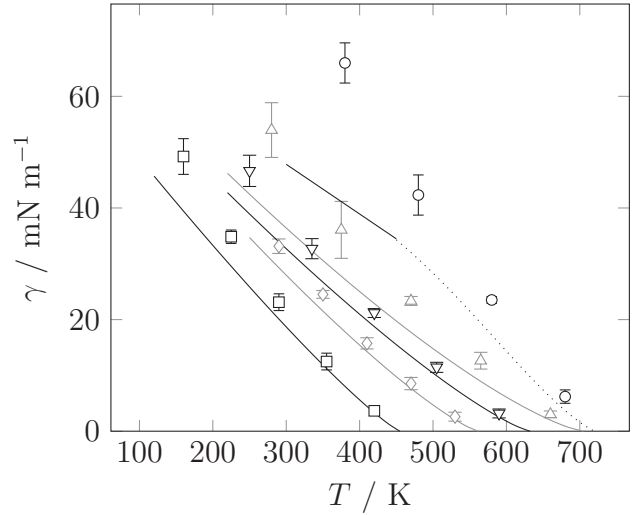}
\caption{Surface tension as a function of the temperature. Solid lines represent DIPPR correlations to experimental data \cite{DIPPR}, the dotted line connects the DIPPR correlation with the critical point of ethylene glycol, and symbols are the present simulation results: Ethylene glycol ($\circ$), ortho-dichlorobenzene ($\triangle$), chlorobenzene ($\triangledown$), benzene ($\diamond$) and phosgene ($\square$).}  
\label{NCLJX_fig:NCLJX5}
\end{figure}

\begin{figure}[h]
\centering
 \includegraphics[width=8.25cm]{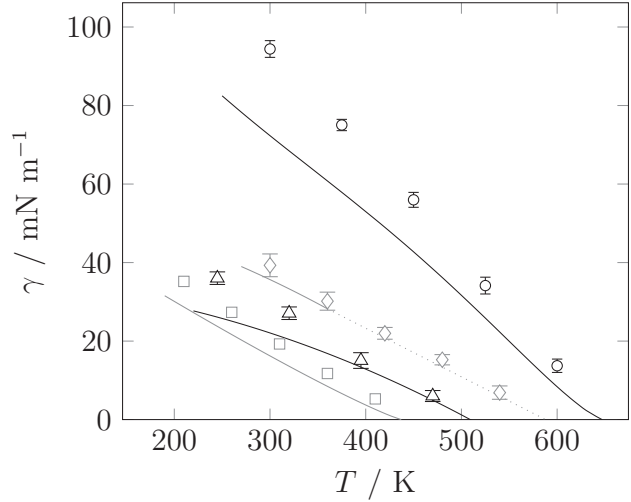}
\caption{Surface tension as a function of the temperature. Solid lines represent DIPPR correlations to experimental data \cite{DIPPR}, the dotted line connects the DIPPR correlation with the critical point of formic acid, and symbols are the present simulation results: Water ($\circ$), formic acid ($\diamond$), methanol ($\triangle$), dimethylamine ($\square$).}  
\label{NCLJX_fig:NCLJX6}
\end{figure}

\begin{figure}[h]
\centering
 \includegraphics[width=8.25cm]{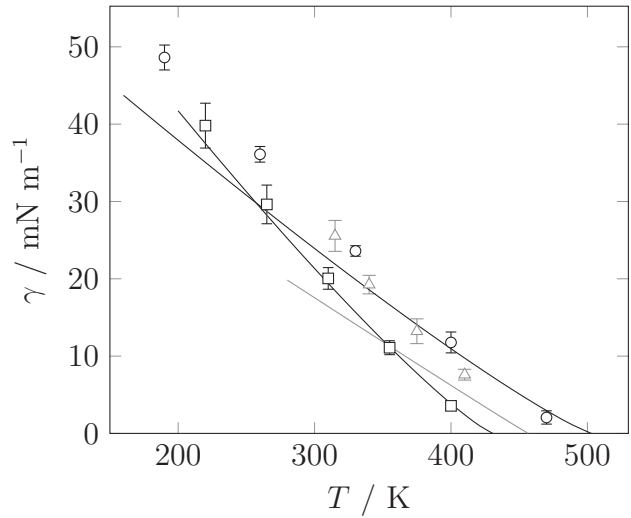}
\caption{Surface tension as a function of the temperature. Solid lines represent DIPPR correlations to experimental data \cite{DIPPR} and symbols are the present simulation results: Dimethyl sulfide ($\circ$), hydrogen cyanide ($\triangle$) and sulfur dioxide ($\square$).}  
\label{NCLJX_fig:NCLJX7}
\end{figure}

\begin{figure}[t]
\centering
 \includegraphics[width=8.25cm]{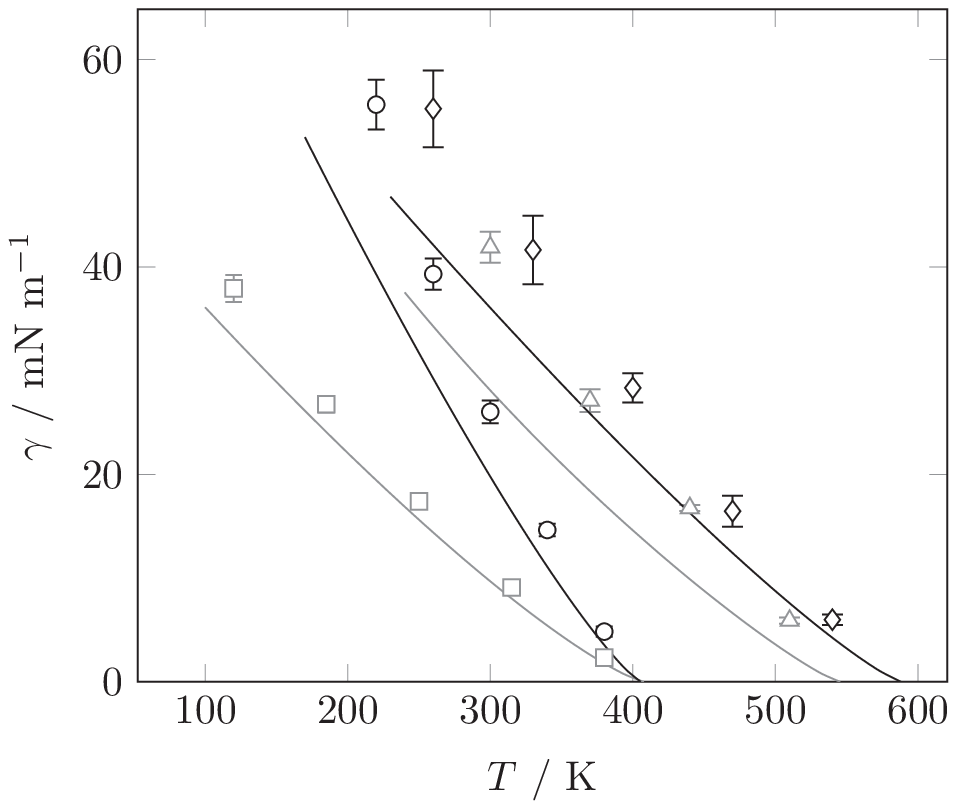}
\caption{Surface tension as a function of the temperature. Solid lines represent DIPPR correlations to experimental data \cite{DIPPR} and symbols are the present simulation results: Nitromethane ($\diamond$), acetonitrile ($\triangle$), ammonia ($\circ$) and isobutane ($\square$).}  
\label{NCLJX_fig:NCLJX8}
\end{figure}

\begin{figure}[t]
\centering
 \includegraphics[width=8.25cm]{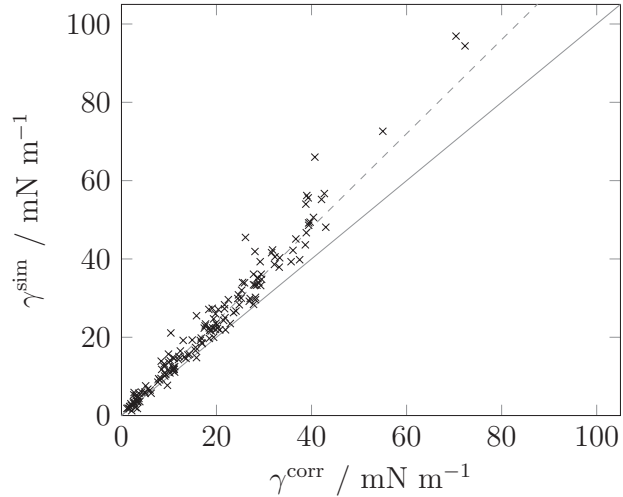}
\caption{Predicted surface tension over the experimental surface tension based on DIPPR correlations \cite{DIPPR} for the present molecular simulation results. The solid line represents perfect agreement between simulation and experiment, and the dashed line represents a deviation of 20 \%.}  
\label{NCLJX_fig:NCLJX_all}
\end{figure}

Figs.~\ref{NCLJX_fig:NCLJX1} -- \ref{NCLJX_fig:NCLJX8} show the surface tension for all studied fluids as a function of the temperature. The molecular simulation results are compared with DIPPR correlations to experimental data. For formaldehyde, methylhydrazine and 1,1-dimethylhydrazine no experimental data are available. The predictions of the surface tension agree reasonably well with the experimental data. Fig.~\ref{NCLJX_fig:NCLJX_all} shows the surface tension predicted by MD simulation as a function of the experimental surface tension calculated by the DIPPR correlation \cite{DIPPR} for all studied fluids. The molecular models overestimate the surface tension in all cases. The average deviation between the predictions by the molecular simulation and the experimental data is about 20 \%. This is in line with results for the surface tension obtained by molecular simulation in the literature \cite{NWLM11,EMRV13,ALGAJM11,HTM15,FLPMM11,EV15,SSDS09,ZLCMSA13,CMHHCS12,WHH16,WSKKHH15}.
\newydd{Compared to other methods for predicting the surface tension of low-molecular fluids,
molecular modeling and simulation, using models which are adjusted to bulk data, leads
to relatively high deviations.}

\newydd{
   Several examples illustrate this: The surface tension of benzene is reproduced with a deviation of
   $\mathnewydd{\left|\delta\gamma\right| \approx 4 \, \%}$ using the corresponding-states (CS)
   correlation by Sastry and Rao \cite{SR95}, $\mathnewydd{\left|\delta\gamma\right| = 1 \, \%}$
   with the CS correlation by Zuo and Stenby \cite{ZS97}, and
   $\mathnewydd{\left|\delta\gamma\right| = 8.5 \, \%}$ from a parachor correlation \cite{HV86,GDDR89,ZS97};
   the molecular model from Huang \textit{et al.}\ \cite{HHHV11} has
   $\mathnewydd{\left|\delta\gamma\right| = 11.9 \, \%}$.
   For cyclohexane, $\mathnewydd{\left|\delta\gamma\right| \approx 1 \, \%}$ is obtained
   following CS by Sastry and Rao \cite{SR95},
   $\mathnewydd{\left|\delta\gamma\right| = 0.7 \, \%}$ with CS by Zuo and Stenby \cite{ZS97},
   $\mathnewydd{\left|\delta\gamma\right| = 7.4 \, \%}$ from the parachor correlation \cite{HV86,GDDR89,ZS97},
   and $\mathnewydd{\left|\delta\gamma\right| = 10.8 \, \%}$ with the molecular model
   from Merker \textit{et al.}\ \cite{MVH12b}.
   In case of ethyl acetate, a CS correlation yields $\mathnewydd{\left|\delta\gamma\right| \approx 7 \, \%}$
   \cite{SR95} and density functional theory with the PC-SAFT equation of state \cite{TG10} reaches
   $\mathnewydd{\left|\delta\gamma\right| \approx 4 \, \%}$, whereas the average relative deviation is
   $\mathnewydd{\left|\delta\gamma\right| = 10.3 \, \%}$ for the molecular model from
   Eckelsbach \textit{et al.}\ \cite{EJKMMV14}
   For methanol, the CS correlation by Sastry and Rao \cite{SR95} exhibits almost
   perfect agreement $\mathnewydd{\left|\delta\gamma\right| < 1 \, \%}$; the molecular model
   by Schnabel \textit{et al.}\ \cite{SSVH07} has $\mathnewydd{\left|\delta\gamma\right| = 35.3 \, \%}$.
   Zuo and Stenby \cite{ZS97} reproduce the surface tension of isobutane with an
   accuracy of $\mathnewydd{\left|\delta\gamma\right| = 1.6 \, \%}$ using their CS correlation
   and with $\mathnewydd{\left|\delta\gamma\right| = 2.5 \, \%}$ using a parachor correlation \cite{HV86,GDDR89,ZS97};
   in contrast, the molecular model for isobutane from Eckl \textit{et al.}\ \cite{EVH08b}
   exhibits a deviation of $\mathnewydd{\left|\delta\gamma\right| = 12.5 \, \%}$.
}

\newydd{
   Square gradient theory with the SAFT-VR Mie equation of
   state, following Garrido \textit{et al.}\ \cite{GMPBM16}, typically
   yields deviations of the order of 2 to 3\% for the surface tension of low-molecular
   fluids. However, no direct comparison is possible with any of the present results.
}

\newydd{
   The anisotropic united atom (AUA) force field \cite{Toxvaerd90, UBDBRF00} was
   also adjusted to bulk data only. All results for $\gamma$ from simulations with
   AUA models are therefore predictions of interfacial properties from bulk fluid properties.
   For benzene, applying the test-area method in Monte Carlo simulations with the
   AUA-9 sites force field, which was parameterized by Nieto Draghi and collaborators \cite{BNU07, NBU07},
   Biscay \textit{et al.}\ \cite{BGLM09} report a surface tension which deviates from
   experimental data by about $\mathnewydd{\delta\gamma \approx +4 \, \%}$, compared to
   $\mathnewydd{\delta\gamma = +11.9 \, \%}$ for the Huang \textit{et al.}\ \cite{HHHV11} model.
   For cyclohexane \cite{BGLM09}, the AUA-9 sites model has $\mathnewydd{\delta\gamma \approx +5 \, \%}$,
   whereas for the Merker \textit{et al.}\ \cite{MVH12b} model,
   $\mathnewydd{\delta\gamma = +11.9 \, \%}$ was found in the present work.
   The AUA-4 model \cite{UBDBRF00} underestimates the surface tension of methanol
   by $\mathnewydd{\delta\gamma \approx -12 \, \%}$, cf.\ Biscay \textit{et al.}\ \cite{BGLM11},
   which compares favorably to the Schnabel \textit{et al.}\ \cite{SSVH07} model with
   $\mathnewydd{\delta\gamma = +35.3 \, \%}$. 
   Overall, the AUA force field is more reliable for predicting the surface tension than the
   models investigated in the present work \cite{BGGLM08}; it has roughly the same accuracy
   as empirical parachor correlations \cite{ZS97}. However, this still makes the AUA force
   field less accurate than empirical CS correlations \cite{SR95, ZS97} and semiempirical
   square gradient theory \cite{GMPBM16}.
}

\newydd{
   Since molecular simulation is also computationally much more expensive than the other
   approaches, it cannot be recommended to predict interfacial properties from molecular models
   which were not previously adjusted to or, at least, validated against such data. However,
}
a systematic overestimation of the surface tension has also been observed in\marked{~(removed: \qq{simulations with})~} density functional theory in combination with physically based equations of state \cite{TT91,FD93,W01,Gross09,LGBJ10}. To account for this overestimation, an empirical correction expression is often employed, which is \newydd{formally} attributed to the presence of capillary waves and decreases the surface tension. Without this correction term, which was adjusted to fit the experimental surface tension values of the n-alkane series \cite{Gross09}, density functional theory would deviate from the surface tension of real fluids in a similar way as the molecular models mentioned above. \newydd{In square gradient theory, the influence parameter, which controls the magnitude of the surface excess free energy, is also adjusted to surface tension data.}
\marked{~(Removed: \qq{Nanoscale roughness of the surface is also present in the molecular simulations. Whether this explains the systematic deviations mentioned above needs further research.})}

\begin{table}[h]
 \caption{Molecular models discussed in the present work. The deviations $\delta \rho'$ and $\delta p^{\rm s}$ as given in respective publications and $\delta\gamma$ is the \newydd{root mean square relative deviation}\marked{~(removed:\qq{relative mean deviation})~} between predictions by the present MD simulations and DIPPR correlations to experimental data \cite{DIPPR}. \newydd{For all considered fluids, the molecular models overestimate $\gamma$ on average.} \# stands for the number of the corresponding sites in the model. \newline}
\centering
\resizebox{0.95 \textwidth}{!}{
 \begin{tabular}{l||l|l|c|c|c|c|l|r|l|l} 
Name &	Formula	& CAS RN & \# LJ & \# Charges & \# Dipoles & \# Quadrupoles & $\delta \rho'$ & $\delta p^{\rm s}$ & $\delta \gamma$ & Source	 \\
\hline
Neon			& Ne	 	 	& 7440-01-9	 & 1 & - & - & - & 0.6 & 2.5 & 20.1 & Vrabec \textit{et al.}~\cite{VSH01} \\
Argon			& Ar	 	 	& 7440-37-1	 & 1 & - & - & - & 0.9 & 2.1 & 31.9 & Vrabec \textit{et al.}~\cite{VSH01} \\
Krypton			& Kr	 	 	& 7439-90-9	 & 1 & - & - & - & 0.9 & 2.1 & 27.8 & Vrabec \textit{et al.}~\cite{VSH01} \\
Xenon			& Xe	 	 	& 7440-63-3	 & 1 & - & - & - & 1.4 & 1.7 & 46.9 & Vrabec \textit{et al.}~\cite{VSH01} \\
Methane			& CH$_4$ 	 	& 74-82-8	 & 1 & - & - & - & 1.0 & 2.7 & 19.2 & Vrabec \textit{et al.}~\cite{VSH01} \\
Acetonitrile		& C$_2$H$_3$N 	 	& 75-05-8	 & 3 & - & 1 & - & 0.1 & 4.7 & 51.7 & Deublein \textit{et al.}~\cite{DMVH12} \\
Cyclohexane		& C$_6$H$_{12}$	 	& 110-82-7	 & 6 & - & - & - & 0.3 & 1.7 & 10.8 & Merker \textit{et al.}~\cite{MVH12b} \\
Cyclohexanone		& C$_6$H$_{10}$O 	& 108-94-1	 & 7 & - & 1 & - & 0.9 & 2.7 & 26.0 & Merker \textit{et al.}~\cite{MVH12b} \\
Cyclohexanol		& C$_6$H$_{10}$OH	& 108-93-0	 & 7 & 3 & - & - & 0.2 & 3.0 & 28.3 & Merker \textit{et al.}~\cite{MVH12} \\
Ethylene oxide		& C$_2$H$_4$O	 	& 75-21-8	 & 3 & - & 1 & - & 0.4 & 1.5 & 16.6 & Eckl \textit{et al.}~\cite{EVH08a} \\
Isobutane		& C$_4$H$_{10}$	 	& 75-28-5	 & 4 & - & 1 & 1 & 0.6 & 4.2 & 12.5 & Eckl \textit{et al.}~\cite{EVH08b} \\
Formaldehyde		& CH$_2$O	 	& 50-00-0	 & 2 & - & 1 & - & 0.9 & 4.3 & $\phantom{1}$-    & Eckl \textit{et al.}~\cite{EVH08b} \\
Dimethyl ether		& C$_2$H$_6$O	 	& 115-10-6	 & 3 & - & 1 & - & 0.4 & 2.6 & 18.9 & Eckl \textit{et al.}~\cite{EVH08b} \\
Sulfur dioxide		& SO$_2$	 	& 7446-09-5	 & 3 & - & 1 & 1 & 0.9 & 4.0 & $\phantom{1}$3.4 & Eckl \textit{et al.}~\cite{EVH08b} \\
Dimethyl sulfide	& C$_2$H$_6$S	 	& 75-18-3	 & 3 & - & 1 & 2 & 0.7 & 4.0 & 18.1 & Eckl \textit{et al.}~\cite{EVH08b} \\
Thiophene		& C$_4$H$_10$	 	& 110-02-1	 & 5 & - & 1 & 1 & 1.2 & 3.8 & 22.4 & Eckl \textit{et al.}~\cite{EVH08b} \\
Hydrogen cyanide	& HCN		 	& 74-90-8	 & 2 & - & 1 & 1 & 1.0 & 7.2 & 51.9 & Eckl \textit{et al.}~\cite{EVH08b} \\
Nitromethane		& CH$_3$NO$_2$	 	& 75-52-5	 & 4 & - & 1 & 1 & 0.2 &18.7 & 31.5 & Eckl \textit{et al.}~\cite{EVH08b} \\
Phosgene		& COCl$_2$	 	& 75-44-5	 & 4 & - & 1 & 1 & 0.5 & 2.1 & 17.2 & Huang \textit{et al.}~\cite{HHHV11} \\
Benzene			& C$_6$H$_6$	 	& 71-43-2	 & 6 & - & - & 6 & 0.4 & 3.4 & 11.9 & Huang \textit{et al.}~\cite{HHHV11} \\
Chlorobenzene		& C$_6$H$_5$Cl	 	& 108-90-7	 & 7 & - & 1 & 5 & 0.9 & 5.0 & 17.8 & Huang \textit{et al.}~\cite{HHHV11} \\
Ortho-Dichlorobenzene	& C$_6$H$_4$Cl$_2$ 	& 95-50-1	 & 8 & - & 1 & 4 & 0.5 & 6.4 & 34.3 & Huang \textit{et al.}~\cite{HHHV11} \\
Cyanogen chloride	& NCCl		 	& 506-77-4	 & 3 & - & 1 & 1 & 0.3 & 2.1 & 15.3 & Miroshnichenko \textit{et al.}~\cite{MGSV13} \\
Cyanogen 		& C$_2$N$_2$	 	& 460-19-5	 & 4 & - & - & 1 & 0.6 &13.0 & $\phantom{1}$2.5 & Miroshnichenko \textit{et al.}~\cite{MGSV13} \\
Heptafluoropropane (R227ea)& C$_3$HF$_7$	& 431-89-0	 &10 & - & 1 & 1 & 1.0 & 1.0 & $\phantom{1}$7.2 & Eckl \textit{et al.}~\cite{EHVH07}  \\
Ammonia			& NH$_3$	 	& 7664-41-7	 & 1 & 4 & - & - & 0.7 & 1.6 & 36.7 & Eckl \textit{et al.}~\cite{EVH08c}  \\
Formic acid		& CH$_2$O$_2$	 	& 64-18-6	 & 3 & 4 & - & - & 0.8 & 5.1 & $\phantom{1}$9.5 & Schnabel \textit{et al.}~\cite{SCVLH07}  \\
Methanol		& CH$_3$OH	 	& 67-56-1	 & 2 & 3 & - & - & 0.6 & 1.1 & 35.3 & Schnabel \textit{et al.}~\cite{SSVH07}  \\
Dimethylamine		& C$_2$H$_7$N	 	& 124-40-3	 & 3 & 3 & - & - & 0.4 & 6.2 & 28.7 & Schnabel \textit{et al.}~\cite{SVH08}  \\
Ethylene glycol		& C$_2$H$_6$O$_2$ 	& 107-21-1	 & 4 & 6 & - & - & 0.8 & 4.8 & 32.6 & Huang \textit{et al.}~\cite{HMHHV12} \\
Water			& H$_2$O	 	& 7732-18-5	 & 1 & 3 & - & - & 1.1 & 7.2 & 30.7 & Huang \textit{et al.}~\cite{HMHHV12} \\
Hydrazine		& N$_2$H$_4$	 	& 302-01-2	 & 2 & 6 & - & - & 0.5 & 7.6 & 29.2 & Elts \textit{et al.}~\cite{EWSV12} \\
Methylhydrazine		& CH$_6$N$_2$	 	& 60-34-4	 & 3 & 3 & - & - & 0.2 & 7.0 & $\phantom{1}$-    & Elts \textit{et al.}~\cite{EWSV12} \\
1,1-Dimethylhydrazine	& C$_2$H$_8$N$_2$ 	& 57-14-7	 & 4 & 3 & - & - & 1.3 & 3.7 & $\phantom{1}$-    & Elts \textit{et al.}~\cite{EWSV12} \\
Ethyl acetate		& C$_4$H$_8$O$_2$ 	& 141-78-6	 & 6 & 5 & - & - & 0.1 & 4.6 & 10.3 & Eckelsbach \textit{et al.}~\cite{EJKMMV14} \\
Decafluorobutane	& C$_4$F$_{10}$ 	& 355-25-9	 &14 &14 & - & - & 0.5 & 3.5 & 10.3 & K\"oster \textit{et al.}~\cite{KNWRV12} \\
Hexamethyldisiloxane	& C$_6$H$_{18}$OSi$_2$	& 107-46-0	 & 9 & 3 & - & - & 0.5 & 5.0 & 12.9 & Thol \textit{et al.}~\cite{TDRWKSV16} \\
Octamethylcyclotetrasiloxane& C$_8$H$_{24}$O$_2$Si$_4$& 556-67-2 &16 & 8 & - & - & 0.5 & 6.0 & 10.5 & Thol \textit{et al.}~\cite{TRKDWSV16} \\
 \end{tabular}
 }
 \label{NCLJX_tab:PredRealFluids}
\end{table}

\newydd{The unfavorable performance of molecular models, compared to methods which are more abstract physically and less expensive numerically, is explained by the fact that the molecular models are \textit{entirely} predictive for interfacial properties. All other methods, i.e.\ density functional theory, square gradient theory, and phenomenological correlations, were adjusted to surface tension data at least indirectly.
Moreover, methods which are based on analytical equations of state, including molecular equations of state, fail in the vicinity of the critical point, so that $\delta\gamma$ diverges at high temperatures. Renormalization group theory has to be employed in these cases to avoid unphysical behavior \cite{TG10, FLVTG11, FLVTG13}.
By molecular simulation, the (Ising class) critical scaling behavior of intermolecular pair potentials is correctly captured. Therefore, any fit of force-field parameters to VLE data over a significant temperature range always indirectly adjusts the molecular model to the critical temperature \cite{SKHKH16}.}

It has been shown before that a better agreement of molecular simulation results with experimental data for the surface tension can be achieved by taking into account experimental data on the surface tension in the parameterization of the molecular models. However,\marked{~(removed: \qq{significant})~} improvements in the quality of the representation of the surface tension\marked{~(removed: \qq{are usually only possible})~} \newydd{have to be traded off against}\marked{~(removed: \qq{when important})~} losses in the quality of the representation of \newydd{bulk fluid properties}\marked{~(removed: \qq{either the vapor pressure or the saturated liquid density are taken into account})} \cite{SKRHKH14,WSKKHH15,SKHKH16,WSHH16}.

\section{Conclusion}

In the present work, the surface tension of real fluids was determined by \newydd{MD}\marked{~(removed: \qq{molecular dynamics})~} simulation. The surface tension was evaluated for 38 real fluids which were parameterized to reproduce the saturated liquid density, vapor pressure and enthalpy of vaporization. The agreement between the calculated bulk values in heterogeneous simulations and the experimental data is very good. On the basis of well described phase equilibria the surface tension was predicted. The surface tension is consistently overpredicted by the molecular models. On average the deviation is about +20 \%. Such overpredictions have been reported before in the literature. The reasons should be investigated in more detail in future research.

\section*{Acknowledgement}

The authors gratefully acknowledge financial support from BMBF within the SkaSim project (grant no. 01H13005A) and from Deutsche Forschungsgemeinschaft (DFG) within the Collaborative Research Center (SFB) 926. They thank Patrick Leidecker for performing some of the present simulations and Gabriela Guevara Carri\'on, Maximilian Kohns, Yonny Mauricio Mu\~noz Mu\~noz and Jadran Vrabec for fruitful discussions. The present work was conducted under the auspices of the Boltzmann-Zuse Society for Computational Molecular Engineering (BZS), and the simulations were carried out on \textit{Elwetritsch} at the Regional University Computing Center Kaiserslautern (RHRK) under the grant TUKL-MSWS and on \textit{SuperMUC} at Leibniz Supercomputing Center (LRZ), Garching, within the SPARLAMPE computing project (pr48te). \marked{\newydd{(Minor corrections were made to this paragraph.)}}

\bibliographystyle{unsrt}

\newpage

\appendix*

\linespread{0.956}
\begin{longtable}{ll|lllll}
 \caption{Molecular simulation results for the vapor-liquid equilibrium of the pure components from the present work. The numbers in parentheses indicate the statistical uncertainties of the last decimal digits.} \\
\hline  \hline 
$T$   &&   $p^\text{s}$  &  $\rho^{'}$	&  $\rho^{''}$	&  $\gamma$ \\
 K    &&    MPa 		& mol l$^{-1}$	&  mol l$^{-1}$ &  mN m$^{-1}$ \\
  \hline
  Acetonitrile &\\
  \hline
 300  && 0.005(4)  & 18.885(1)  & 0.002(1)  &  41.9(15) \\ 
 370  && 0.137(38) & 17.013(20) & 0.053(12) &  27.1(6)  \\ 
 440  && 0.839(90) & 14.810(21) & 0.307(58) &  15.7(30) \\
 510  && 2.72(14)  & 11.84(9)   & 1.09(15)  &  $\phantom{1}$5.9(3)  \\
  \hline  
  Cyclohexane &\\
  \hline
 280  && 0.008(6)  & $\phantom{1}$9.355(3)  & 0.003(1)  &  29.2(12) \\ 
 335  && 0.049(8)  & $\phantom{1}$8.750(4)  & 0.017(3)  &  22.3(13)  \\ 
 390  && 0.270(33) & $\phantom{1}$8.093(3)  & 0.090(9)  &  15.5(10) \\
 445  && 0.855(34) & $\phantom{1}$7.325(13) & 0.273(11) &  $\phantom{1}$9.5(7)  \\
 510  && 2.387(93) & $\phantom{1}$6.117(26) & 0.834(41) &  $\phantom{1}$3.5(4)  \\
  \hline 
  Cyclohexanone &\\
  \hline
 250  && 0.000(0)  & $\phantom{1}$9.902(8)  & 0.000(0)  &  50.6(38) \\ 
 315  && 0.002(2)  & $\phantom{1}$9.327(4)  & 0.001(1)  &  40.5(12)  \\ 
 380  && 0.020(6)  & $\phantom{1}$8.741(5)  & 0.006(1)  &  30.8(21) \\
 445  && 0.148(22) & $\phantom{1}$8.118(11) & 0.042(5)  &  22.3(13)  \\
 510  && 0.558(37) & $\phantom{1}$7.423(14) & 0.146(9)  &  14.8(6)  \\
  \hline 
  Cyclohexanol &\\
  \hline
 300  && 0.000(0)  & $\phantom{1}$9.700(22) & 0.000(0)  &  40.3(62) \\ 
 375  && 0.009(7)  & $\phantom{1}$9.028(4)  & 0.003(1)  &  32.0(13)  \\ 
 450  && 0.122(20) & $\phantom{1}$8.277(3)  & 0.034(8)  &  22.9(4) \\
 525  && 0.633(24) & $\phantom{1}$7.422(13) & 0.165(12) &  14.0(7)  \\
 600  && 2.093(91) & $\phantom{1}$6.325(20) & 0.575(31) &  $\phantom{1}$5.7(11)  \\
  \hline
  Ethylene oxide &\\
  \hline
 180  && 0.000(0)  & 23.230(4)  & 0.000(0)  &  48.1(33) \\ 
 245  && 0.006(4)  & 21.325(6)  & 0.004(3)  &  38.6(23)  \\ 
 310  && 0.331(68) & 19.323(27) & 0.145(30) &  25.0(13) \\
 375  && 1.304(57) & 16.851(13) & 0.495(27) &  14.3(19)  \\
 440  && 4.77(40)  & 13.39(13)  & 2.16(34)  &  $\phantom{1}$3.7(11)  \\
  \hline  
  Isobutane &\\
  \hline
 120  && 0.000(0)  & 12.654(6)  & 0.000(0)  &  37.9(13) \\ 
 185  && 0.001(1)  & 11.539(3)  & 0.001(1)  &  26.7(8)  \\ 
 250  && 0.067(17) & 10.383(23) & 0.032(7)  &  17.4(5)  \\
 315  && 0.584(10) & $\phantom{1}$8.983(28) & 0.263(16) &  $\phantom{1}$9.1(7) \\
 380  && 2.32(12)  & $\phantom{1}$7.112(41) & 1.13(12)  &  $\phantom{1}$2.3(4) \\
  \hline  
  Formaldehyde &\\
  \hline
 180  && 0.001(1)  & 31.984(54) & 0.001(1)  &  51.6(14) \\ 
 235  && 0.030(9)  & 29.228(5)  & 0.017(5)  &  36.3(15) \\ 
 290  && 0.331(54) & 26.222(77) & 0.155(20) &  24.2(14) \\
 345  && 1.66(13)  & 22.594(39) & 0.782(66) &  12.7(9) \\
\hline  \hline
\pagebreak \hline \hline
$T$   &&   $p^\text{s}$  &  $\rho^{'}$	&  $\rho^{''}$	&  $\gamma$ \\
 K    &&    MPa 		& mol l$^{-1}$	&  mol l$^{-1}$ &  mN m$^{-1}$ \\
  \hline  
  Dimethyl ether &\\
  \hline
 205  && 0.016(8)  & 17.016(5)  & 0.016(4)  &  30.0(12) \\ 
 260  && 0.205(39) & 15.415(53) & 0.100(35) &  19.8(13) \\ 
 315  && 0.95(13)  & 13.563(13) & 0.442(38) &  10.9(10) \\
 370  && 3.34(35)  & 10.94(12)  & 1.70(31)  &  $\phantom{1}$3.6(6) \\
  \hline
  Sulfur dioxide & \\
  \hline
 220  && 0.013(8)  & 24.771(11) & 0.007(4)  &  39.8(29) \\ 
 265  && 0.106(39) & 22.923(25) & 0.049(14) &  29.6(27)  \\ 
 310  && 0.65(17)  & 20.932(52) & 0.278(74) &  20.0(14) \\
 355  && 1.95(46)  & 18.48(12)  & 0.83(19)  &  11.1(9)  \\
 400  && 4.65(26)  & 14.87(28)  & 2.30(25)  &  $\phantom{1}$3.6(10)  \\
  \hline  
  Dimethyl sulfide & \\
  \hline
 190  && 0.001(1)  & 19.996(5)  & 0.000(0)  &  48.6(16) \\ 
 260  && 0.036(14) & 18.300(17) & 0.022(14) &  36.1(10) \\ 
 330  && 0.342(26) & 16.471(18) & 0.134(16) &  23.6(7) \\
 400  && 1.70(15)  & 14.315(56) & 0.635(65) &  11.8(13)  \\
 470  && 5.32(33)  & 10.85(26)  & 2.60(57)  &  $\phantom{1}$2.1(8)  \\
  \hline  
  Thiophene &\\
  \hline
 250  && 0.002(1)  & 13.107(7)  & 0.001(1)  &  43.6(22) \\ 
 320  && 0.035(16) & 12.168(6)  & 0.013(4 ) &  33.3(7)  \\ 
 390  && 0.263(37) & 11.166(7)  & 0.084(9)  &  22.4(13) \\
 460  && 1.12(11)  & 10.001(16) & 0.342(35) &  13.6(10) \\
 530  && 3.13(29)  & $\phantom{1}$8.458(65)  & 1.01(16)  &  $\phantom{1}$5.0(8)  \\
  \hline  
  Hydrogen cyanide &\\
  \hline
 280  && 0.038(21) & 25.756(38)  & 0.021(11)  &  34.0(11) \\ 
 315  && 0.122(12) & 24.015(12)  & 0.063(10)  &  25.5(21)  \\ 
 340  && 0.343(32) & 22.589(23)  & 0.153(23)  &  19.2(12) \\
 375  && 0.949(50) & 20.50(12)   & 0.431(41)  &  13.2(17) \\
 410  && 2.03(10)  & 17.766(69)  & 1.00(12)   &  $\phantom{1}$7.6(7)  \\
  \hline
  Nitromethane &\\
  \hline
 260  && 0.001(1)  & 19.569(16) & 0.001(1)  &  55.3(37) \\ 
 330  && 0.030(7)  & 18.014(6)  & 0.014(6)  &  41.6(34) \\ 
 400  && 0.109(22) & 16.259(52) & 0.044(15) &  28.3(14) \\
 470  && 0.826(69) & 14.337(25) & 0.266(32) &  16.5(15) \\
 540  && 2.84(26)  & 11.70(20)  & 1.06(15)  &  $\phantom{1}$6.0(9)  \\
  \hline  
  Phosgene &\\
  \hline
 160  && 0.000(0)  & 17.047(11) & 0.000(0)  &  49.2(32) \\ 
 225  && 0.007(7)  & 15.578(10) & 0.004(3)  &  34.8(17) \\ 
 290  && 0.133(34) & 14.047(7)  & 0.058(8)  &  23.1(21) \\
 355  && 0.937(36) & 12.311(20) & 0.367(8)  &  12.5(12) \\
 420  && 3.35(33)  & $\phantom{1}$9.873(48) &  1.46(22)  &  $\phantom{1}$3.7(7)  \\
  \hline  \hline
\pagebreak \hline \hline
$T$   &&   $p^\text{s}$  &  $\rho^{'}$	&  $\rho^{''}$	&  $\gamma$ \\
 K    &&    MPa 		& mol l$^{-1}$	&  mol l$^{-1}$ &  mN m$^{-1}$ \\
  \hline 
  Benzene &\\
  \hline
 290  && 0.007(7)  & 11.208(13) & 0.003(3)  &  33.2(13) \\ 
 350  && 0.092(23) & 10.403(7)  & 0.033(3)  &  24.3(7)  \\ 
 410  && 0.423(47) & $\phantom{1}$9.523(3)  & 0.137(16) &  15.7(10) \\
 470  && 1.441(67) & $\phantom{1}$8.450(19) & 0.468(22) &  $\phantom{1}$8.5(11) \\
 530  && 3.337(92) & $\phantom{1}$6.884(36) & 1.21(13)  &  $\phantom{1}$2.6(9)  \\
  \hline 
  Chlorobenzene &\\
  \hline
 250  && 0.004(3)  & 10.320(8)  & 0.002(2)  & 46.7(28) \\ 
 335  && 0.011(6)  & $\phantom{1}$9.466(11) & 0.005(3)  &  33.3(19)\\ 
 420  && 0.122(13) & $\phantom{1}$8.567(23) & 0.037(2)  &  21.9(8) \\
 505  && 0.79(14)  & $\phantom{1}$7.520(30) & 0.222(44) &  11.7(9) \\
 590  && 2.91(8)   & $\phantom{1}$5.992(20) & 0.94(7)   &  $\phantom{1}$3.2(9)  \\
  \hline
  Ortho-Dichlorobenzene & \\
  \hline
 375  && 0.003(2)  & $\phantom{1}$8.472(5) & 0.001(1)  &  36.1(51) \\ 
 470  && 0.097(17) & $\phantom{1}$7.631(3) & 0.026(6)  &  23.4(8)  \\ 
 565  && 0.701(89) & $\phantom{1}$6.647(16)& 0.173(28) &  12.6(15) \\
 660  && 2.79(14)  & $\phantom{1}$5.240(33)& 0.807(80) &  $\phantom{1}$3.0(7) \\
  \hline  
  Cyanogen chloride & \\
  \hline
 280  && 0.079(21) & 19.777(13) & 0.036(8)  &  28.3(21) \\ 
 315  && 0.312(36) & 18.490(46) & 0.135(9)  &  21.6(16)  \\ 
 350  && 0.86(17)  & 17.08(39)  & 0.361(83) &  14.9(21) \\
 385  && 1.86(43)  & 15.27(7)   & 0.79(23)  &  $\phantom{1}$9.4(17) \\
 420  && 3.74(36)  & 12.92(9)   & 1.86(54)  &  $\phantom{1}$4.2(7) \\
  \hline  
  Cyanogen & \\
  \hline
 260  && 0.094(31) & 18.030(33) & 0.049(21) &  21.9(24) \\ 
 290  && 0.349(95) & 16.866(58) & 0.168(25) &  16.6(12)  \\ 
 320  && 0.79(30)  & 15.546(99) & 0.36(17)  &  11.5(16) \\
 350  && 2.17(34)  & 13.820(16) & 1.04(26)  &  $\phantom{1}$5.7(6) \\
 380  && 4.40(5)   & 11.15(28)  & 2.89(40)  &  $\phantom{1}$1.2(6) \\
  \hline 
  Heptafluoropropane & \\
  \hline
 200  && 0.002(1)  & 10.364(13) & 0.001(1)  &  21.9(10) \\ 
 250  && 0.063(10) & $\phantom{1}$9.381(8)  & 0.031(2)  &  14.8(19)  \\ 
 300  && 0.466(39) & $\phantom{1}$8.215(3)  & 0.213(21) &  $\phantom{1}$7.7(9) \\
 350  && 1.776(68) & $\phantom{1}$6.531(20) & 0.938(46) &  $\phantom{1}$1.8(6) \\
  \hline
  Ammonia &  \\
  \hline
 220  && 0.029(2)  & 41.985(19) & 0.015(1)  &  55.1(19) \\ 
 260  && 0.233(38) & 38.927(14) & 0.116(23) &  40.1(36)  \\ 
 300  && 1.029(80) & 35.470(30) & 0.467(39) &  26.2(13) \\
 340  && 3.080(66) & 31.410(26) & 1.421(43) &  15.2(25) \\
 380  && 7.23(15)  & 25.572(21) & 3.86(21)  &  $\phantom{1}$4.8(1) \\
  \hline   \hline
\pagebreak \hline \hline
$T$   &&   $p^\text{s}$  &  $\rho^{'}$	&  $\rho^{''}$	&  $\gamma$ \\
 K    &&    MPa 		& mol l$^{-1}$	&  mol l$^{-1}$ &  mN m$^{-1}$ \\
  \hline  
  Formic acid & \\
  \hline
 300  && $\phantom{1}$0.014(10) & 26.192(5)  & 0.007(2)  &  39.3(29) \\ 
 360  && $\phantom{1}$0.072(9)  & 24.478(9)  & 0.038(14) &  30.2(23)  \\ 
 420  && $\phantom{1}$0.333(28) & 22.607(11) & 0.169(20) &  22.0(15) \\
 480  && $\phantom{1}$1.14(17)  & 20.36(17)  & 0.543(61) &  15.3(13) \\
 540  && $\phantom{1}$3.20(25)  & 17.45(14)  & 1.57(18)  &  $\phantom{1}$6.7(15) \\
  \hline  
  Methanol & \\
  \hline
 245  && $\phantom{1}$0.001(1)  & 26.175(9)  & 0.001(1)  &  36.0(16) \\ 
 320  && $\phantom{1}$0.045(17) & 23.958(6)  & 0.020(7)  &  27.1(16) \\ 
 395  && $\phantom{1}$0.58(13)  & 21.317(36) & 0.359(49) &  14.0(20) \\
 470  && $\phantom{1}$3.61(14)  & 17.318(91) & 1.61(26)  &  $\phantom{1}$6.0(14) \\
  \hline  
  Dimethylamine & \\
  \hline
 210  && $\phantom{1}$0.003(1)  & 16.653(10) & 0.002(1)  &  35.2(13) \\ 
 260  && $\phantom{1}$0.033(11) & 15.523(4)  & 0.016(6)  &  27.3(4) \\ 
 310  && $\phantom{1}$0.257(24) & 14.279(7)  & 0.109(10) &  19.0(2) \\
 360  && $\phantom{1}$1.04(10)  & 12.888(6)  & 0.407(45) &  11.8(9) \\
 410  && $\phantom{1}$2.65(74)  & 10.814(77) & 1.07(50)  &  $\phantom{1}$5.3(18) \\
  \hline
  Ethylene glycol & \\
  \hline
 380  && $\phantom{1}$0.005(3)  & 16.867(29) & 0.002(1)  &  65.6(21) \\ 
 480  && $\phantom{1}$0.107(60) & 15.527(25) & 0.028(11) &  42.3(36) \\ 
 580  && $\phantom{1}$1.144(52) & 13.759(14) & 0.268(9)  &  23.6(31) \\
 680  && $\phantom{1}$5.45(58)  & 10.906(97) & 1.44(19)  &  $\phantom{1}$6.2(12) \\
  \hline   
  Water & \\
  \hline
 300  && $\phantom{1}$0.006(5)  & 56.348(12) & 0.002(1)  &  94.0(21) \\ 
 375  && $\phantom{1}$0.069(16) & 52.650(12) & 0.024(8)  &  75.1(13) \\ 
 450  && $\phantom{1}$0.63(10)  & 48.475(11) & 0.31(13)  &  56.9(19) \\
 525  && $\phantom{1}$3.31(16)  & 43.412(11) & 0.94(6)   &  34.1(21) \\
 600  && 10.43(20) & 36.485(46) & 3.28(3)   &  13.7(20) \\
  \hline    
  Hydrazine & \\
  \hline
 280  && $\phantom{1}$0.000(0)  & 32.321(8)  & 0.000(0)  &  96.7(23) \\ 
 360  && $\phantom{1}$0.028(10) & 29.955(5)  & 0.010(5)  &  72.6(13) \\ 
 440  && $\phantom{1}$0.358(24) & 27.220(18) & 0.153(19) &  49.4(30) \\
 520  && $\phantom{1}$2.24(64)  & 24.063(80) & 1.16(33)  &  29.9(20) \\
  \hline    
  Methylhydrazine & \\
  \hline
 270  && $\phantom{1}$0.009(5)  & 19.384(10) & 0.004(3)  &  57.3(13) \\ 
 345  && $\phantom{1}$0.031(13) & 17.944(6)  & 0.012(3)  &  43.1(16) \\ 
 420  && $\phantom{1}$0.420(64) & 16.358(48) & 0.137(40) &  29.7(12) \\
 495  && $\phantom{1}$1.93(21)  & 14.598(16) & 0.573(51) &  16.7(8)  \\
 570  && $\phantom{1}$5.21(22)  & 12.21(12)  & 1.73(13)  &  $\phantom{1}$5.7(9) \\
  \hline \hline
\pagebreak \hline \hline
$T$   &&   $p^\text{s}$  &  $\rho^{'}$	&  $\rho^{''}$	&  $\gamma$ \\
 K    &&    MPa 		& mol l$^{-1}$	&  mol l$^{-1}$ &  mN m$^{-1}$ \\
  \hline    
  1,1-Dimethylhydrazine & \\
  \hline
 260  && 0.003(2)  & 13.783(6)  & 0.001(1)  &  35.5(21) \\ 
 315  && 0.045(13) & 12.833(4)  & 0.018(6)  &  27.6(16) \\ 
 370  && 0.290(14) & 11.806(6)  & 0.101(9)  &  19.4(9)  \\
 425  && 1.03(13)  & 10.603(3)  & 0.344(40) &  11.9(4)  \\
 480  && 2.97(19)  & $\phantom{1}$9.026(33)  & 1.10(12)  &  $\phantom{1}$4.6(4) \\
  \hline    
  Ethyl acetate & \\
  \hline
 190  && 0.000(0) & 11.726(4)  & 0.000(0)  &  42.2(19) \\ 
 265  && 0.002(1) & 10.709(18) & 0.001(1)  &  29.9(6) \\ 
 340  && 0.063(12)& $\phantom{1}$9.636(10) & 0.024(3) &  19.7(9)  \\
 415  && 0.497(39)& $\phantom{1}$8.379(6)  & 0.170(6) &  10.9(12)  \\
 490  && 2.078(53)& $\phantom{1}$6.584(19) & 0.831(37)&  $\phantom{1}$3.0(7) \\
  \hline   
  Decafluorobutane & \\
  \hline
 260  && 0.077(13)& $\phantom{1}$6.848(12) & 0.038(7)  &  11.8(17) \\ 
 310  && 0.357(6) & $\phantom{1}$6.021(28) & 0.157(5)  &  $\phantom{1}$6.5(6) \\ 
 360  && 1.329(67)& $\phantom{1}$4.852(77) & 0.657(43) &  $\phantom{1}$1.7(3)  \\
  \hline    
  Hexamethyldisiloxane & \\
  \hline
 210  && 0.000(0) & $\phantom{1}$5.211(31) & 0.000(0)  &  26.3(17) \\ 
 280  && 0.002(1) & $\phantom{1}$4.788(5)  & 0.001(0)  &  18.4(9) \\ 
 350  && 0.047(15)& $\phantom{1}$4.327(8)  & 0.017(4)  &  12.1(11)  \\
 420  && 0.299(20)& $\phantom{1}$3.771(10) & 0.097(10) &  $\phantom{1}$6.7(4) \\ 
 490  && 1.161(26)& $\phantom{1}$3.020(25) & 0.421(10) &  $\phantom{1}$1.8(2) \\
  \hline  
  Octamethylcyclotetrasiloxane & \\
  \hline
 310  && 0.000(0) & $\phantom{1}$3.135(14) & 0.000(0)  &  18.6(33) \\ 
 355  && 0.004(2) & $\phantom{1}$2.970(12) & 0.001(0)  &  14.6(10) \\ 
 420  && 0.044(14)& $\phantom{1}$2.709(8)  & 0.013(4)  &  10.1(11)  \\
 485  && 0.217(15)& $\phantom{1}$2.403(12) & 0.061(4)  &  $\phantom{1}$5.7(2) \\ 
 550  && 0.699(12)& $\phantom{1}$1.985(27) & 0.211(7)  &  $\phantom{1}$1.9(3)  \\
  \hline    \hline
\label{NCLJX_tab:data}
\end{longtable}

\section*{Molecular simulation details}

The equation of motion was solved by a leapfrog integrator~\cite{Fincham92} with a time step of $\Delta t$ = 1 fs. The elongation of the simulation volume normal to the interface was 30 nm and the thickness of the liquid film in the center of the simulation volume was 15 nm to account for finite size effects~\cite{WLHH13}. The elongation in the other spatial directions was at least 10 nm. 
The equilibration was executed for 500,000 time steps. The production was conducted for 2,500,000 time steps to reduce statistical uncertainties. The statistical errors were estimated to be three times the standard deviation of five block averages, each over 500,000 time steps. The saturated densities and vapor pressures were calculated as an average over the respective phases excluding the area close to the interface, i.e.\ the area where the first derivative of the density with respect to the $y$ coordinate deviated from zero significantly. 

The cutoff radius was set to 17.5 \AA~and a center-of-mass cutoff scheme was employed. The Lennard-Jones interactions were corrected with a slab-based long range correction based on the density profile \cite{WRVHH14}. Electrostatic long-range interactions were approximated by a resulting effective molecular dipole and corrected with a slab-based long range correction based on the density profile \cite{WHH15}. The quadrupolar interactions do not need a long-range correction as they decay by $r^{-10}$\newydd{, cf.\ Prausnitz \textit{et al.}}\ \cite{PLA98}


\marked{
   (Removed Appendix B, previously consisting of Fig.\ 12 and the following text: \qq{Based on a correlation from previous work \cite{WLHH13}, the surface tension of the one-center Lennard-Jones fluids can be calculated \cite{VSH01}. The results for noble gases and methane are shown in Fig.~[previously 12].})
} 


\marked{
   (Removed a figure, i.e.\ originally Fig.\ 12, from the manuscript.)
}

\end{document}